\documentstyle[12pt]{article}

\let\ph=\varphi  \let\PH=\Phi

\def\0{\over } \def\1{\vec }     \def\2{{1\over2}} \def\4{{1\over4}}
\def\5{\bar }  \def\6{\partial } \def\7#1{{#1}\llap{/}}
\def\8#1{{\textstyle{#1}}}       \def\9#1{{\bf {#1}}}
 \def\llp{\hbox to 0pt{\hss /\hskip1.5pt}}
\def\llo{\hbox to 0.2pt{\hss /}} \def\llq{\hbox to 0pt{\hss /\hskip0.5pt}}
\def\so{\supset\hbox to 0pt{\hss $\displaystyle -$\hskip1pt}}

\def\<{\langle } \def\>{\rangle }

\let\nn=\nonumber
\def\bea{\begin{eqnarray}} \def\eea{\end{eqnarray}}
\def\beann{\begin{eqnarray*}} \def\eeann{\end{eqnarray*}}
\def\beq{\begin{equation}} \def\eeq{\end{equation}}

\date{}
\title{
{\large\rm DESY 94-043}\hfill{\large\tt ISSN 0418-9833}\\
{\large\rm March 1994}\hfill\vspace*{3.5cm}\\
Gauge Invariant Treatment \\
of the Electroweak Phase Transition}
\author{W. Buchm\"{u}ller, Z. Fodor\thanks{On leave from Institute for
Theoretical Physics, E\"otv\"os University, Budapest, Hungary}\hspace{.3cm}and
A. Hebecker\\
{\normalsize\it Deutsches Elektronen-Synchrotron DESY, Hamburg D-22603,
Germany}\vspace*{3.5cm}\\
}
\begin{document}

\setlength{\baselineskip}{18pt}
\maketitle
\begin{abstract}
We evaluate the gauge invariant effective potential for the composite field
$\sigma=2\PH^{\dagger}\PH$ in the SU(2)-Higgs model
at finite temperature. Symmetric and broken
phases correspond to the domains $\sigma\leq T^2/3$ and $\sigma > T^2/3$,
respectively. The effective potential
increases very steeply at small values of $\sigma$. Predictions
for several observables, derived from the ordinary and the gauge
invariant effective potential, are compared. Good agreement is found
for the critical temperature and the jump in the order parameter.
The results for the latent heat differ significantly for large Higgs masses.
\end{abstract}
\newpage
Detailed recent studies of the electroweak phase transition are all based
on the effective action for the Higgs field $\PH$ in Landau gauge
[1--4].
Although the effective action is gauge dependent, physical observables,
derived from this action, must be gauge independent. To
verify this explicitly is an important and non-trivial task, especially
for quantities like the surface tension,
which involve not only the effective potential
but also derivative terms.

Hence, it appears desirable to use a manifestly gauge invariant approach
as far as possible. This is of particular importance for comparison
with lattice Monte Carlo simulations which are usually carried out without
gauge fixing \cite{lattice}. It is known that the expectation value of the
operator $\PH^{\dagger}\PH$ is well suited to characterize the broken or Higgs
phase in lattice simulations \cite{J}, and the corresponding effective
potential has been evaluated \cite{L}.

In the following we will calculate the effective potential for the composite
field $\sigma=2\PH^{\dagger}\PH$ in continuum perturbation theory. It turns out
that this potential, which is gauge invariant by definition,
has two qualitatively new features. First,
the symmetric phase, which in the conventional framework corresponds
to a single point, $\PH=0$, is now related to the half-axis $\sigma<T^2/3$.
The local minimum in this domain is rather narrow, and at small values
of $\sigma$ the potential increases very steeply.
Second, the new potential is valid at temperatures above and below the
critical temperature $T_c$ and can in fact be smoothly extrapolated down
to $T=0$.

For simplicity, we shall restrict our discussion in this paper to the
Higgs model in three dimensions. With proper identification of parameters
we will then obtain the finite-temperature result for small values of the
Higgs field.
We have also performed the analogous calculation at finite temperature,
which will be discussed, together with two-loop results,
in a forthcoming paper \cite{bufohe}.

The SU(2)-Higgs model in three dimensions is described by
the lagrangian
\beq
\cal{L}={1\over 4}W^a_{\mu\nu}W^a_{\mu\nu}+(D_{\mu}\PH)^{\dagger}
D_{\mu}\PH + V_0(\ph^2)\quad,
\eeq
where
\beq
V_0(\ph^2)={1\over 2}m^2 \ph^2 + {1\over 4}\mu\lambda \ph^4\ ,
\ \ph^2=2\PH^{\dagger}\PH\ .
\eeq
Here $W^a_{\mu\nu}$ is the ordinary field strength tensor and
$D_{\mu}=\partial_{\mu}-i\mu^{1/2}{g\over 2} W^a_{\mu} \tau^a$
is the covariant derivative; $g$ and $\lambda$ are gauge and scalar
self-coupling, respectively, and $\mu$ is the mass scale used to define
dimensionless couplings in three dimensions.

In order to obtain the effective potential for the field
$\sigma=2\PH^{\dagger}\PH$, one has to evaluate first the ``free energy''
in the presence of an external source $J$,
\beq
e^{\textstyle -\Omega W(J)}=\int DW D\PH D\PH^{\dagger}
e^{\textstyle -S(\PH,\PH^{\dagger},W)-\int d^3x 2\PH^{\dagger}\PH J}\ ,
\eeq
where $\Omega$ is the total volume.
For constant $J$ the effective potential is obtained as the Legendre transform,
\bea \label{legendre}
{\partial \over \partial J}W(J) &=& \sigma\ ,\nn \\
V(\sigma) &=& W(J(\sigma))-\sigma J\ .
\eea

The free energy $W(J)$ can be calculated in the standard
semiclassical or loopwise
expansion \footnote{See, e.g., ref.\ \cite{zinn}, chapter 5.3}.
The equation for a spatially constant stationary point,
\beq
(m^2 + 2\mu\lambda \PH_c^{\dagger}\PH_c+2J)\PH_c = 0\ ,
\eeq
has two solutions, $\PH_c=\PH_s$ and $\PH_c=\PH_b$, which correspond
to the symmetric and the broken phase, respectively,
\bea
\PH_s &=& 0\ ,\\
\PH_b &=& \left(-{1\over 2\mu\lambda} (m^2+2J)\right)^{1/2}\PH_0\ ,
\eea
where $|\PH_0|=1$.
The determinants of fluctuations around the two stationary points
depend on the masses of vector bosons, Higgs ($\ph$) and Goldstone ($\chi$)
bosons. In the
broken phase ($\PH_c=\PH_b$) one has, in any covariant gauge,
\beq
m^2_W = -{g^2(m^2 + 2J)\over 4\lambda}\ ,
\ m^2_{\ph} = -2(m^2 + 2J)\ ,\  m^2_{\chi} = 0\ ,
\eeq
whereas in the symmetric phase ($\PH_c=\PH_s$) the masses are given by
\beq
m^2_W = 0\ , \ m^2_{\ph} = m^2_{\chi} = m^2 + 2J\ .
\eeq
$\PH_s$ and $\PH_b$ correspond to the global minima of the classical action
for $m^2+2J>0$ and $m^2+2J<0$, respectively.

The one-loop contribution to the free energy in covariant gauge (cf.
\cite{DJ}) is given by
\bea \label{oneloop}
W_1(J) = {1\over 2}\int {d^3k\over (2\pi)^3}
\Big( &&\!\!\!\!\!\!\!\!\!6\ln{(k^2 + m_W^2)} + \ln{(k^2 + m_{\ph}^2)}
\nn \\ &&\!\!\!\!\!\!\!\!\!
+ 3\ln{(k^4 + k^2m_{\chi}^2+\alpha m_W^2m_\chi^2)}-6\ln{k^2}\,\Big)\ ,
\eea
where $\alpha$ is the gauge parameter. This expression is gauge independent
since the product $m_Wm_\chi$ vanishes in the symmetric and broken phase. One
easily verifies that the same result for $W_1(J)$ is obtained in $R_\xi$-gauge.

Subtracting linear divergencies by means of
dimensional regularization one obtains for the finite part,
\beq
W(J) = W_b(J) \Theta(-m^2-2J) + W_s(J) \Theta(m^2+2J)\ ,
\eeq
where
\bea
W_b(J) &=& -{1\over 4\mu\lambda}(m^2+2J)^2 - {1\over 2\pi}\left(-
{g^2\over 4\lambda}(m^2+2J)\right)^{3/2} \\
& &\quad -{1\over 12\pi}\left(-2(m^2+2J)\right)^{3/2}\ ,\\
W_s(J) &=& -{1\over 3\pi}\left(m^2+2J\right)^{3/2}\ .
\eea
Note, that in the broken phase only  vector bosons and the Higgs boson
contribute.
In the symmetric phase all four scalar degrees of freedom
contribute equally, whereas the gauge boson contribution vanishes.

The one-loop effective potential can now be obtained by performing
the Legendre transformation according to eq.\ (\ref{legendre}).
In the broken phase, to order $\cal{O}(g^3,\lambda^{3/2})$,
it is sufficient to determine $\sigma(J)$ from
the tree-level expression for $W(J)$. The resulting potential, which is not
convex, represents the energy of an appropriately defined, homogeneous state
\cite{WW}. A straightforward calculation yields
\beq \label{vgi}
V(\sigma) = V_b(\sigma) \Theta(\sigma) + V_s(\sigma) \Theta(-\sigma)\ ,
\eeq
where
\bea
V_b(\sigma) &=& {1\over 2} m^2 \sigma + {1\over 4}\mu \lambda \sigma^2
- {1\over 2\pi}\left({1\over 4}\mu g^2\sigma\right)^{3/2}
- {1\over 12\pi}\left(2\mu\lambda \sigma\right)^{3/2}\ ,
\label{broken}\\
V_s(\sigma) &=& {1\over 2} m^2 \sigma - {\pi^2\over 6}\sigma^3\ .
\label{symm}
\eea
Here the couplings depend on the renormalization scale, i.e.,
$g=g(\mu)$, $\lambda=\lambda(\mu)$, $m^2=m^2(\mu)$.

It is very instructive to compare this result with the familiar
effective potential for the field $\PH$ in Landau gauge. The one-loop
correction is given by the integral in eq.\ (\ref{oneloop}).
Shifting the scalar field $\PH$ in the usual way by the
background field $\ph/\sqrt{2}$ yields the masses,
\beq
m_W^2={1\over 4}\mu g^2\ph^2\ ,
\ m_{\ph}^2=m^2 + 3\mu\lambda\ph^2\ ,
\ m_{\chi}^2=m^2 + \mu\lambda\ph^2\ .
\eeq
For the potential in Landau gauge one then obtains
\beq \label{vlg}
V_{LG}(\ph^2) = {1\over 2} m^2 \ph^2 + {1\over 4}\mu \lambda \ph^4
- {1\over 12\pi}\left(6 m_W^3 + m_{\ph}^3 + 3 m_{\chi}^3 \right)\ .
\eeq

Comparing the two potentials (\ref{vgi}) and (\ref{vlg}) the first
striking difference is the range of the fields. For the potential
(\ref{vlg}) one has $0\leq \ph^2< \infty$, whereas for the potential
(\ref{vgi}) the field varies in the range $-\infty <\sigma<
\infty$. In the first case the symmetric phase is represented by the
point $\ph =0$, whereas in the second case it corresponds to the
half-axis $\sigma\leq 0$. This difference is a consequence of
the different source terms for which the ``free energy'' $W$ is
calculated.
Note, that for the gauge invariant potential
at one-loop order  only the four scalar degrees
of freedom contribute in the symmetric phase. At small values of $\sigma$
the potential increases very steeply.

The second important difference between the potentials (\ref{vgi}) and
(\ref{vlg}) concerns the contribution of scalar loops in the broken
phase. Contrary to the ordinary potential, the non-analytic terms
of the gauge invariant potential do not depend on $m^2$.
Hence, this potential can also be used for $m^2<0$, where the symmetric phase
is unstable.

We are interested in the SU(2)-Higgs model at finite temperature, for
which the ordinary, ring-improved one-loop potential is given by
(cf.\ \cite{bufo})
\bea \label{vtlg}
V_{ring}(\ph^2,T) &=& {1\over 2}\left({3\over 16}g^2 + {1\over 2}\lambda
\right)\left(T^2-T^2_b\right)\ph^2 + {1\over 4}\lambda\ph^4 \\
& &- {T\over 12\pi}\left( 3m_L^3 + 6m_T^3 + m_{\ph}^3 + 3m_{\chi}^3\right)
+ \cal{O}(g^4, \lambda^2).
\eea
The masses of longitudinal and transverse vector bosons,
Higgs and Goldstone bosons are
\bea
m_L^2 &=& {5\over 6}g^2T^2 + {1\over 4}g^2\ph^2\ ,\
m_T^2 = {1\over 4}g^2\ph^2\ ,\\
m_{\ph}^2 &=& \left({3\over 16}g^2 + {1\over 2}\lambda\right)(T^2-T_b^2)
+ 3\lambda \ph^2\ ,\\
m_{\chi}^2 &=& \left({3\over 16}g^2 + {1\over 2}\lambda\right)(T^2-T_b^2)
+ \lambda \ph^2\ .
\eea
Here we have neglected the effect of the top quark, which is not of interest
for our discussion. For values of the Higgs field $\ph$ small compared
to the temperature $T$, one can expand $m_L$ in powers of $\ph^2/T^2$. Up to
terms of order $\cal{O}(g^3\ph^6/T^2)$ the result is then identical
with the effective potential of the three-dimensional theory in Landau
gauge (cf. \cite{JP}),
\beq
V_{ring}(\ph^2,T) = T V_{LG}\left({\ph^2\over T}\right)\ ,
\eeq
if parameters are identified as follows,
\bea
\mu=T\ ,\ \lambda(\mu)=\lambda-{3\over 128 \pi}\sqrt{{6\over 5}}g^3\ ,
\ g(\mu)=g\ ,\hphantom{xxxxxxxxxxx}\\
m^2(\mu)=\left({3\over 16}g^2 + {1\over 2}\lambda
-{3\over 16\pi}\sqrt{{5\over 6}}g^3\right)(T^2-\bar{T}_b^2)\ ,
\ \bar{T}_b^2=\frac{3g^2 + 8\lambda}{3g^2+8\lambda-{{3\over\pi}}
\sqrt{{5\over 6}}g^3}
\ .
\eea
Inserting these parameters into eqs.\ (\ref{vgi}) - (\ref{symm})
we finally obtain for
the gauge invariant finite-temperature potential,
\beq \label{vtgi}
V(\sigma,T) = V_s(\sigma,T)\Theta(\sigma) + V_b(\sigma,T)\Theta(-\sigma)\ ,
\eeq
where
\bea
V_b(\sigma,T) &=& {1\over 2}m^2(T)\sigma + {1\over 4}\lambda\sigma^2
- {T\over 12\pi}\left(6\left({1\over 4}g^2\sigma\right)^{3/2}
+ \left(2\lambda\sigma\right)^{3/2}\right)
\ ,\\
V_s(\sigma,T) &=& {1\over 2}m^2(T)\sigma - {\pi^2\over 6}{\sigma^3\over T^2}\ .
\eea
Contrary to the conventional potential $V_{ring}(\ph^2,T)$, the gauge
invariant potential $V(\sigma,T)$ is valid at temperatures above and below the
barrier temperature $\bar{T}_b$.

We have also evaluated the gauge invariant effective potential directly in the
finite-temperature theory \cite{bufohe}. In the high temperature expansion
the result is essentially the same, it can be obtained from eq.\ (\ref{vtgi})
by a shift in the field $\sigma$,
\beq
\sigma \rightarrow \sigma - {1\over 3}T^2\ .
\eeq
This shift is obtained by subtracting divergencies using dimensional
regularization. Note, that in general the shift is arbitrary and fixed by
a renormalization condition. Only its temperature dependence has physical
significance.

In figs.\ (1) and (2) the ordinary potential $V_{ring}(\ph^2,T)$ and
the gauge invariant potential $V(\sigma,T)$ are shown for a Higgs mass of
$m_H = 70$ GeV. Each potential is shown at its critical temperature,
defined by the degeneracy of the two minima. In the broken
phase the two potentials are very similar. The main difference concerns
the symmetric phase where the gauge invariant potential shows a strong
increase at small values of $\sigma$. Note, that the barrier is higher
by about a factor of two for the gauge invariant potential.

We have evaluated several observables for the two potentials. The
critical temperatures are different, but very similar. For Higgs
masses between 30 GeV and 120 GeV the ratio $(T_c-\bar{T}_b)/\bar{T}_b$
differs by at most 40\%. Fig.\ (3) shows that
also the predictions for $\ph_c$, the shift
in the Higgs field at the critical temperature, are in good agreement.
The latent heat, shown in fig.\ (4), differs by about 70\% at
$m_H=120$ GeV. We expect similar discrepancies for the surface tension.

In summary, the main differences between the ordinary effective potential
in Landau gauge and the new gauge invariant potential concern the
symmetric phase and the effect of scalar loops in the broken phase.
The new potential can be used at temperatures above and below the
barrier temperature. The gauge invariance of the new potential
should be an advantageous
feature also with respect to non-perturbative
effects which are expected to be important in the symmetric phase
of the non-abelian, and possibly also the abelian, Higgs model.

We gratefully acknowledge helpful discussions with K. Jansen, M. L\"uscher,
\linebreak I. Montvay and M. Reuter.

\vspace{.7cm}
\noindent
{\bf\Large Figure captions}\\[.5cm]
{\bf Fig.1} Effective potential in Landau gauge, $m_H=70$ GeV.\\
{\bf Fig.2} The gauge invariant effective potential, $m_H=70$ GeV.\\
{\bf Fig.3} \parbox[t]{14.5cm}{\setlength{\baselineskip}{18pt}
Shift of the Higgs field as function of the
Higgs mass, as predicted by the potential in Landau gauge (dashed line) and
the gauge invariant potential (full line).}\\[.2cm]
{\bf Fig.4} \parbox[t]{14.5cm}{\setlength{\baselineskip}{18pt}
Latent heat as function of the Higgs mass,
as predicted by the potential in Landau gauge (dashed line) and
the gauge invariant potential (full line).}
\end{document}